\documentclass{moriondNOTITLES}
\usepackage{physics} % for \dd, \dv derivative
\usepackage{siunitx} % for \SI
\usepackage{xspace} % for \xspace
\usepackage{xcolor} % for \color
\usepackage{multirow} % for \multirow (\mro)

\makeatletter % to get reference number without superscript
\newcommand{\inlinecite}[1]{\nocite{#1}\csname b@#1\endcsname}
\makeatother

\newcommand{\Photo}{\includegraphics[height=35mm]{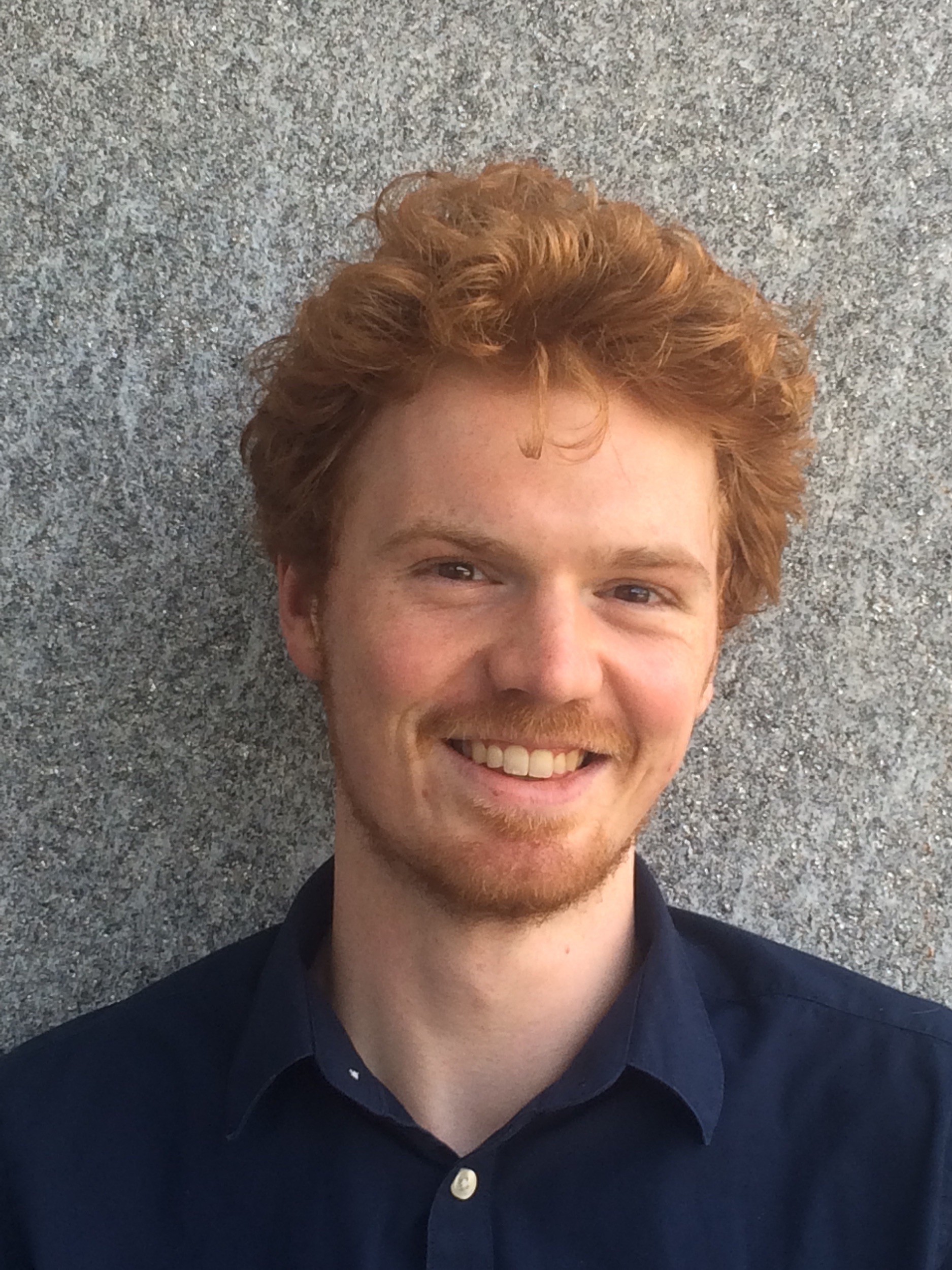}}
\colorlet{myred}{red!80!black}
\colorlet{myblue}{blue!85!black}
\def\mco{\multicolumn}
\def\mro{\multirow}
\newcommand\A[1]{{\color{myblue}\hypersetup{citecolor=myblue}A\,\cite{ATLAS-#1}}} % ATLAS
\newcommand\C[1]{{\color{myred}\hypersetup{citecolor=myred}C\,\cite{CMS-#1}}} % CMS
\def\NA{---}

\newcommand{\pt}{\ensuremath{p_\text{T}}\xspace}
\newcommand{\ptmiss}{\ensuremath{p_\text{T}^\text{miss}}\xspace}
\newcommand{\PQc}{\ensuremath{\text{c}}\xspace}
\newcommand{\PQb}{\ensuremath{\text{b}}\xspace}
\newcommand{\BF}{\mathcal{B}}
\newcommand{\LQ}{\ensuremath{\text{LQ}}\xspace}
\newcommand{\mLQ}{\ensuremath{m_{\text{LQ}}}\xspace}
\newcommand{\mll}{\ensuremath{m_{\ell\ell}}\xspace}
\newcommand{\mmm}{\ensuremath{m_{\mu\mu}}\xspace}

\newcommand{\mthth}{\ensuremath{m_{\tau_\text{had}\tau_\text{had}}}\xspace}
\newcommand{\mcc}{\ensuremath{m_\text{cc}}\xspace}

\newcommand{\thetastar}{\ensuremath{\theta^*}\xspace}
\newcommand{\costhetastar}{\ensuremath{\cos\theta^*}\xspace}
\newcommand{\cstar}{\ensuremath{c_*}\xspace}
\begin{document}

%%%%%%%%%%%%%
%   TITLE   %
%%%%%%%%%%%%%

\vspace*{4cm}
\title{Leptoquark searches at ATLAS and CMS}
\author{I. Neutelings on behalf of the ATLAS and CMS Collaborations}
\address{CERN, Switzerland}

%%%%%%%%%%%%%%%%
%   ABSTRACT   %
%%%%%%%%%%%%%%%%

\maketitle
\abstracts{An overview of searches for leptoquarks in proton-proton collisions at ATLAS and CMS are presented, with a focus on results from the Run-2 dataset collected between 2016 and 2018 at $\sqrt{s}=\SI{13}{TeV}$, and corresponding to an integrated luminosity of 137--\SI{140}{fb^{-1}}. In particular, recent results are highlighted.}

%%%%%%%%%%%%%%%%%%%%
%   INTRODUCTION   %
%%%%%%%%%%%%%%%%%%%%

\section{Introduction}

% THEORY
One of the most striking features of the Standard Model (SM) is its neat organization of quarks and leptons into three distinct generations. %, which seem to be exact copies of each other
Each generation of fermions seems to be a copy of the others, except that the experimentally measured fermion masses vary by many orders of magnitudes.
Nevertheless, the coupling strength between SM gauge bosons and leptons is identical across flavors, a property referred to as lepton flavor universality (LFU).
%The pattern does not appear natural or accidental, but instead very hierarchical.
%This could point to some yet undiscovered physics that preferentially couples to third-generation fermions.
This symmetry appears to be accidental, as the SM provides no theoretical principle that enforces it.

% EXPERIMENT
Any violation of LFU in Nature would be a clear indication of new physics beyond the SM (BSM).
In fact, over the past two decades, experimental evidence that is in tension with the SM has emerged, pointing in this direction. This includes anomalies in rare semileptonic decays of B mesons~\cite{Marin,Eklund,Reboud,Martinov}---some of which have been presented at this conference---as well as the precise measurement of the muon's anomalous magnetic moment ($g-2$)~\cite{muon_g-2}.

% LEPTOQUARKS
These anomalies can be explained by a new hypothetical type of boson, called the leptoquark (LQ), which may simultaneously couple to leptons and quarks. Such particles arise naturally in many theories beyond the SM, such as Grand Unified Theories, Supersymmetry with $R$-parity violation, compositeness, and technicolor. Their coupling  structure to quarks and lepton may \emph{a priori} vary between flavors, explicitly violating LFU and potentially accounting for the observed anomalies mentioned above.

% THIS NOTE
This note will give an overview of searches for leptoquarks in ATLAS~\cite{ATLAS}, CMS~\cite{CMS} using the Run-2 dataset collected between 2016 and 2018 at a center-of-mass energy of $\sqrt{s}=\SI{13}{TeV}$, and corresponding to an integrated luminosity of 137--\SI{140}{fb^{-1}}.
In particular, the most recent results are highlighted.

%%%%%%%%%%%%%%%%%%%
%   LEPTOQUARKS   %
%%%%%%%%%%%%%%%%%%%

%\section{LQ models \& production at the LHC}
\section{Overview of LQ searches at the LHC} %in Run-2 data}
\label{sec:overview}

LQs are either scalar or vector bosons that can couple simultaneously to a quark and a lepton. They therefore carry both lepton and baryon number, as well as color charge and a fractional electric charge. Searches at the LHC typically use simplified models with the following model parameters:
\begin{itemize}
  \item LQ mass \mLQ, ranging roughly between 0.2 and \SI{10}{TeV}.
  \item Coupling strength $\lambda$ (sometimes $g$ for vector LQ) to quarks and leptons, ranging roughly between 0 and 4.
  \item Branching fraction $\beta=\BF(\LQ\to q\ell)=1-\BF(\LQ\to q'\nu)$ to a final state with charged leptons.
  \item In the case of vector LQs, ``non-minimal'' coupling $\kappa$ to gluons (or sometimes other gauge bosons as well), such that $\kappa=0$ corresponds to a minimal coupling, while $\kappa=1$ corresponds to a Yang--Mills scenario.
\end{itemize}
Exact conventions of definition or notation may vary between publications.
As most experimental search focus on a particular final state with or without neutrinos, they focus on benchmark scenarios such as $\beta=1$ (no neutrinos), $\beta=1$ (a mix), $\beta=0$ (no charged leptons), %(no LQ couplings to neutrinos), $\beta=0.5$ (an equal mix of charged leptons and neutrino) or $\beta=0$ (no LQ couplings to charged leptons),
although by combining different final states, one can set lower exclusion limits on $\mLQ$ as a function of $\beta$.

% FIGURE: Feynman diagrams
% https://cms-results.web.cern.ch/cms-results/public-results/publications/EXO-19-016/
\begin{figure}[b!]
  \centerline{
  \raisebox{-0.5\height}{%
    \includegraphics[page=10,width=0.26\textwidth]{fig/LQ_colored.pdf}%
  }\hspace{0.5mm}
  \raisebox{-0.5\height}{%
    \includegraphics[page=1,width=0.26\textwidth]{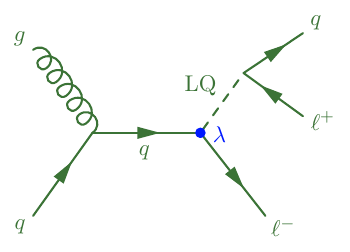}%
  }\hspace{0.5mm}
  \raisebox{-0.5\height}{%
    \includegraphics[page=8,width=0.22\textwidth]{fig/LQ_colored.pdf}%
  }\hspace{0.5mm}
  \raisebox{-0.5\height}{%
    \includegraphics[page=13,width=0.20\textwidth]{fig/LQ_colored.pdf}%
  }}
  \vspace{-3mm}
  \caption{%
Example Feynman diagrams of the main LQ production mechanisms at the LHC. From left to right: LQ pair, gluon-induced single LQ, photon-induced single LQ production, as well as nonresonant production of two leptons via $t$-channel LQ exchange. The coupling strength between the LQ and fermions is indicated as $\lambda$.%
  }
  \label{fig:diagrams}
  \vspace{-4mm}
\end{figure}

There are several ways LQs can be produced in proton-proton collisions, such as at the LHC. Several representative diagrams are shown in Fig.~\ref{fig:diagrams}.
\begin{itemize}
  \item \textbf{LQ pair production} tends to have a large cross section thanks to its production through QCD, similar to $t\overline{t}$ production. It is largely model-independent as its dominant production mechanisms do not depend on $\lambda$.
  \item \textbf{Single LQ production} through either a gluon- or lepton-induced process. This process grows with $\lambda^2$, as the cross section tends to scale with $\lambda^2$.
  \item \textbf{Nonresonant dilepton production} though $t$-channel LQ exchange. At high $\lambda$, the cross section scales with $\lambda^2/\mLQ^4$.
\end{itemize}
Historically, most experimental searches have focused on pair production because of its large cross section and large kinematics, and their signatures are frequently covered by searches for the pair production of supersymmetric particles.
However, it is important to note that the final states of the three different production modes are not always completely separable, as they typically involve at least two high-\pt leptons (assuming lepton number conservation).
Furthermore, owing to their different dependencies on $\mLQ$ and $\lambda$, searches for different production modes offer strong complementarity, especially at high $\lambda$. %Searches that include single and/or nonresonant LQ production focussing at high $\lambda$ as well.

The first searches for a nonresonant dilepton signature mode have only been published in the past several years. Its nonresonant signature makes it more challenging and it may have a nontrivial interference with the SM processes such as Drell--Yan, however at high enough \mLQ, the process acts as a pure four-fermion interaction and its kinematics becomes largely independent of \mLQ and $\lambda$. This opens up a vast phase space at high \mLQ and $\lambda$.

%%%%%%%%%%%%%%%%
%   OVERVIEW   %
%%%%%%%%%%%%%%%%

%\section{Overview of LQ searches in Run-2 data}
% https://twiki.cern.ch/twiki/bin/view/CMSPublic/SummaryPlotsEXO13TeV#Leptoquark_summary_plot
% https://atlaspo.cern.ch/public/summary_plots/
% https://atlas.web.cern.ch/Atlas/GROUPS/PHYSICS/PUBNOTES/ATL-PHYS-PUB-2024-012/
% https://atlas.web.cern.ch/Atlas/GROUPS/PHYSICS/PUBNOTES/ATL-PHYS-PUB-2025-013/

An overview of leptoquark searches using Run-2 data by the ATLAS ({\color{myblue}A}) and CMS ({\color{myred}C}) Collaborations is shown in Tables~\ref{tab:pair}--\ref{tab:nonres}.
They are ordered by production mechanism and final state to indicate what experimental signatures have. The $j$ symbol denotes a jet from a light-flavored quark (u, d, s), while $\nu$ can be a neutrino of any flavor. Note that Ref.~\inlinecite{CMS-EXO-22-018} is currently only search the photon-induced single production.
As these tables show, most Run-2 searches have focused on final states with third generation fermions, as constraints from the B anomalies favor stronger couplings between LQs and higher-generation fermions. Nevertheless, many opportunities remain such as finals states containing an electron or muon accompanied by a light- or charm-flavored jets.
Summaries of LQ mass exclusions can be found in Refs~\inlinecite{ATLAS_LQ_summary} and \inlinecite{CMS_LQ_summary}.

The following section will highlight recent results from Refs.~\inlinecite{ATLAS-SUSY-2018-25}, \inlinecite{CMS-EXO-22-013}, and \inlinecite{ATLAS-EXOT-2022-42}.

% TABLE: Pair production
\begin{table}[tb]
  \vspace{-4mm}
  % TABLE: Pair production: beta = 0, 1
  \begin{minipage}[t]{0.55\textwidth}
  \centering
  \caption{Searches for the production of a LQ pair at ATLAS ({\color{myblue}A}) and CMS ({\color{myred}C}) with exclusively neutrinos ($\beta=0$) or exclusively charged leptons ($\beta=1$) in the final state.}
  %, ordered by final states, where $j$ stands for light quark flavors (u, d, s).}
  \label{tab:pair}
  \vspace{2mm}
  \renewcommand{\arraystretch}{1.1} % increase line height
  \begin{tabular}{ccccc}
  \hline %---------------------------
             & $jj$ & cc & bb & tt \\
  \hline %---------------------------
  $\nu\nu$   & \C{SUS-19-005}
                    & \A{SUSY-2018-25}, \C{SUS-19-005}
                         & \A{SUSY-2018-34}, \C{SUS-19-005}
                              & \A{SUSY-2018-12}, \C{SUS-19-005} \\
  ee         & \NA %\mro{2}{*}{\NA}
                    & \mro{2}{*}{\A{EXOT-2019-13}}
                         & \A{EXOT-2019-13}
                              & \mro{2}{*}{\A{EXOT-2019-19,ATLAS-EXOT-2020-08}, \C{EXO-21-002}} \\
  $\mu\mu$   & \NA  &
                         & \A{EXOT-2019-13}, \C{EXO-21-019}
                              &   \\
  $\tau\tau$ & \mco{2}{c}{\A{EXOT-2020-18}}
                         & \A{SUSY-2019-18,ATLAS-EXOT-2021-15}, \C{EXO-19-016}
                              & \A{EXOT-2019-15}, \C{EXO-21-002} \\
  \hline %---------------------------
  \end{tabular}
  \end{minipage}\hfill
  % TABLE: Pair production: beta = 0.5
  \begin{minipage}[t]{0.40\textwidth}
  \centering
  \caption{Searches for the production of a LQ pair at ATLAS ({\color{myblue}A}) and CMS ({\color{myred}C}) with mixed final states ($0<\beta<1$).}
  \label{tab:pair_mix}
  \vspace{2mm}
  \renewcommand{\arraystretch}{1.1} % increase line height
  \begin{tabular}{cccc}
  \hline %---------------------------
            & $jj$ & cs & tb \\
  \hline %---------------------------
  e$\nu$    & \NA & \NA & \mro{2}{*}{\A{EXOT-2019-12}} \\
  $\mu\nu$  & \NA & \NA & \\
  $\tau\nu$ & \NA & \NA & \A{SUSY-2019-18}, \C{EXO-19-015} \\
  \hline %---------------------------
  \end{tabular}
  \end{minipage}
  \vspace{-2mm}
\end{table}

% TABLES: Single + nonres. production
\begin{table}[tb]
  %\vspace{-3mm}
  % TABLE: Single production: beta = 0, 1
  \begin{minipage}[t]{0.54\textwidth}
  \centering
  \caption{Searches for single LQ production at ATLAS ({\color{myblue}A}) and CMS ({\color{myred}C}). Parentheses indicate quarks or leptons produced in association that may or may not be explicitly selected for.}
  \label{tab:single}
  \vspace{2mm}
  \renewcommand{\arraystretch}{1.1} % increase line height
  \begin{tabular}{ccccc}
  \hline %----------------------------------
                & $j$ & c   & b(b)(b) & t \\
  \hline %----------------------------------
  $\nu(\nu)$    & \C{EXO-20-004}
                      & \NA & \NA & \NA \\
  e(e)          & \NA & \NA & \NA & \NA \\
  e$\mu$        & \mco{2}{c}{\A{EXOT-2018-29}}
                            & \NA & \NA \\
  $\mu(\mu)$    & \NA & \NA & \NA & \NA \\
  $\tau(\nu)$   & \NA & \NA & \NA & \C{EXO-19-015} \\
  $\tau(\mu q)$ & \NA & \NA & \NA & \A{TOPQ-2023-23}, \C{TOP-22-011} \\
  $\tau(\tau)$  & \C{EXO-22-018}
                      & \NA & \A{EXOT-2022-39}, \C{EXO-19-016,CMS-EXO-22-018}
                                  & \NA \\
  \hline %----------------------------------
  \end{tabular}
  \end{minipage}\hfill
  % TABLE: Nonres. production
  \begin{minipage}[t]{0.41\textwidth}
  \centering
  \caption{Searches for the nonresonant dilepton production through $t$-channel exchange of a LQ at ATLAS ({\color{myblue}A}) and CMS ({\color{myred}C}).}
  \label{tab:nonres}
  \vspace{2mm}
  \renewcommand{\arraystretch}{1.1} % increase line height
  \begin{tabular}{cccc}
  \hline %--------------------------------
           & e   & $\mu$ & $\tau$(b)(b) \\
  \hline %--------------------------------
  $\nu$    & \NA & \NA   & \C{EXO-21-009} \\
  e        & \C{EXO-22-013}
                 & \NA   & \NA \\
  $\mu$    & \NA & \C{EXO-22-013}
                         & \A{TOPQ-2023-23}, \C{TOP-22-011} \\
  $\tau$   & \NA & \NA   & \A{EXOT-2022-39,ATLAS-EXOT-2022-42}, \C{EXO-19-016,CMS-HIG-21-001} \\
  \hline %--------------------------------
  \end{tabular}
  \end{minipage}
  \vspace{-3mm}
\end{table}

%%%%%%%%%%%%%%%%%%%
%   NEW RESULTS   %
%%%%%%%%%%%%%%%%%%%

\section{Highlights of recent results}

% https://atlas.web.cern.ch/Atlas/GROUPS/PHYSICS/PAPERS/SUSY-2018-25/
% https://cms-results.web.cern.ch/cms-results/public-results/publications/SUS-19-005/
\subsection{Search for \texorpdfstring{$\LQ\overline{\LQ}\to \PQc\overline{\PQc}\nu\overline{\nu}$}{LQ LQ -> ccnunu}}
The search in Ref.~\inlinecite{ATLAS-SUSY-2018-25} by the ATLAS Collaboration targets BSM models that contribute to final states with charmed-tagged jets and missing transverse momentum. This includes the pair production of charm or top squarks, as well as the pair production of LQs that decay to a charm quark and neutrino.

\begin{figure}[tb!]
  \vspace{-4mm}
  \centering
  \includegraphics[width=0.46\textwidth]{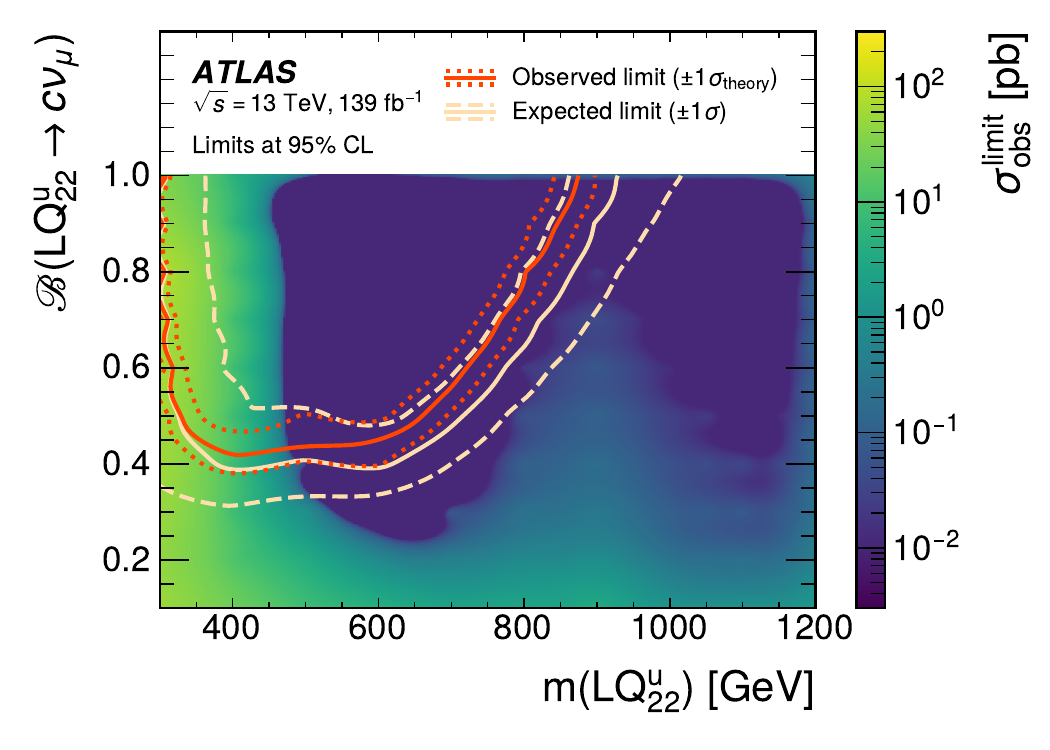}%
  \hspace{1mm}%
  \includegraphics[width=0.45\textwidth]{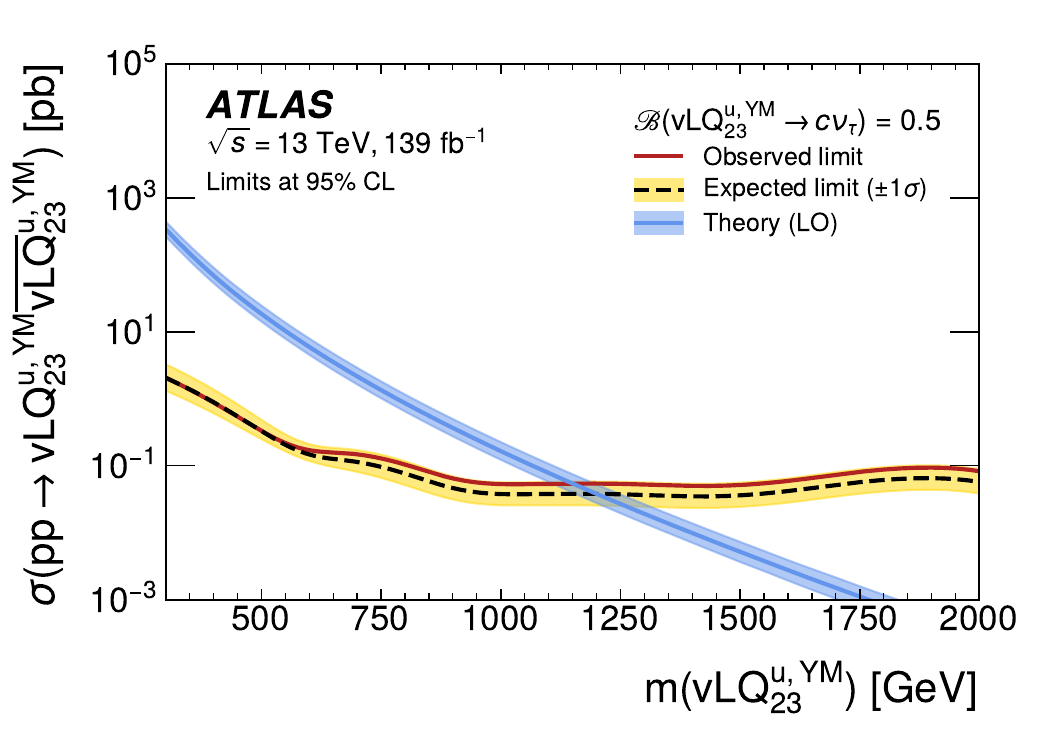}%
  \vspace{-1mm}
  \caption{
Exclusion limits for up-type scalar LQs coupled to second-generation leptons (left) and  $U(1)$ vector LQs with $\kappa=1$ (right), from the search for $\LQ\overline{\LQ}\to \PQc\overline{\PQc} \nu\overline{\nu}$ in Ref.~\protect\inlinecite{ATLAS-SUSY-2018-25}.
  }
  \label{fig:limits_ccnunu}
\end{figure}

% SELECTION
The LQ signal is target by selecting events with high missing momentum ($\ptmiss>\SI{250}{GeV}$), at least two charmed-tagged jets with large \pt and an invariant mass $\mcc>\SI{200}{GeV}$, and no additional electron, muon, $\tau$ lepton, or b-tagged jets.
% SIGNAL REGION
The signal is extracted from three signal regions that are created from cuts on \mcc and \ptmiss significance ($\ptmiss/\sigma[\ptmiss]$), while controlling backgrounds in regions with one or two leptons.
% RESULTS
No significant excess of events over the SM background expectation is observed, and upper limits are set on the productions cross section of several LQ models and coupling scenarios. LQ with masses up to 0.9 (1.5) \unit{TeV} are excluded for a scalar (vector) LQ, as shown in Fig.~\ref{fig:limits_ccnunu}.
% RESULTS
These results can be compared to previous results in Ref.~\inlinecite{CMS-SUS-19-005} by the CMS experiment, which did not exploit charm tagging, but targeted jets flavors inclusively, excluding b jets.
Although that search could distinguish between jet flavors of a potential signal, it set more stringent exclusion limits, as searches for pair production are typically statistically limited.

% FIGURE: ccnunu limits (ATL-SUSY-2018-25)
% https://atlas.web.cern.ch/Atlas/GROUPS/PHYSICS/PAPERS/SUSY-2018-25/

% FIGURE: Nonres. dilep. limits (CMS-EXO-22-013)
% https://cms-results.web.cern.ch/cms-results/public-results/publications/EXO-19-016/
\begin{figure}[b!]
  \centerline{
  \raisebox{-0.5\height}{%
    \includegraphics[width=0.38\textwidth]{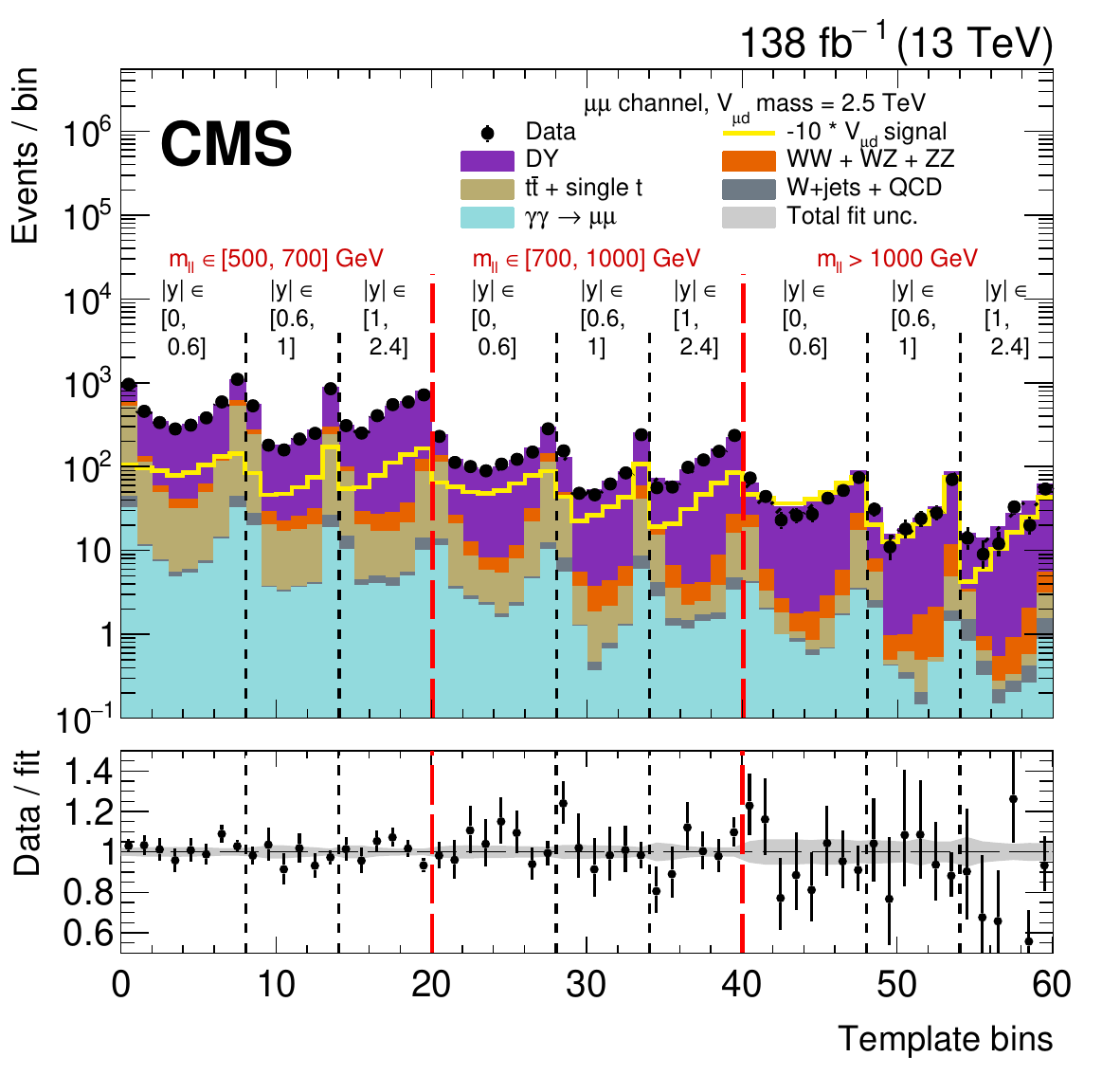}%
  }%\hspace{1pt}%
  \raisebox{-0.5\height}{%
    \includegraphics[width=0.32\textwidth]{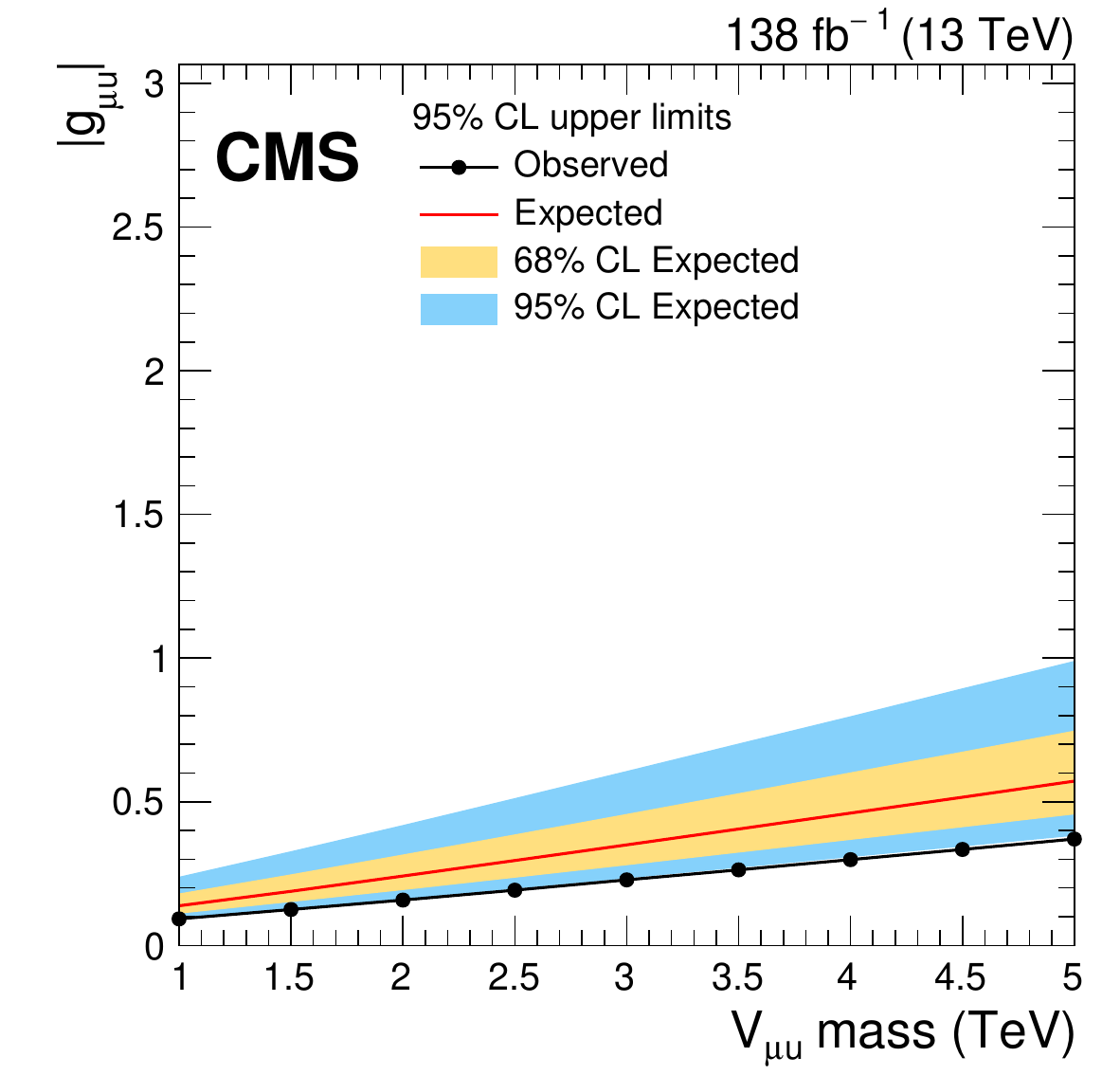}%
  }%\hspace{1pt}%
  \raisebox{-0.5\height}{%
    \includegraphics[width=0.32\textwidth]{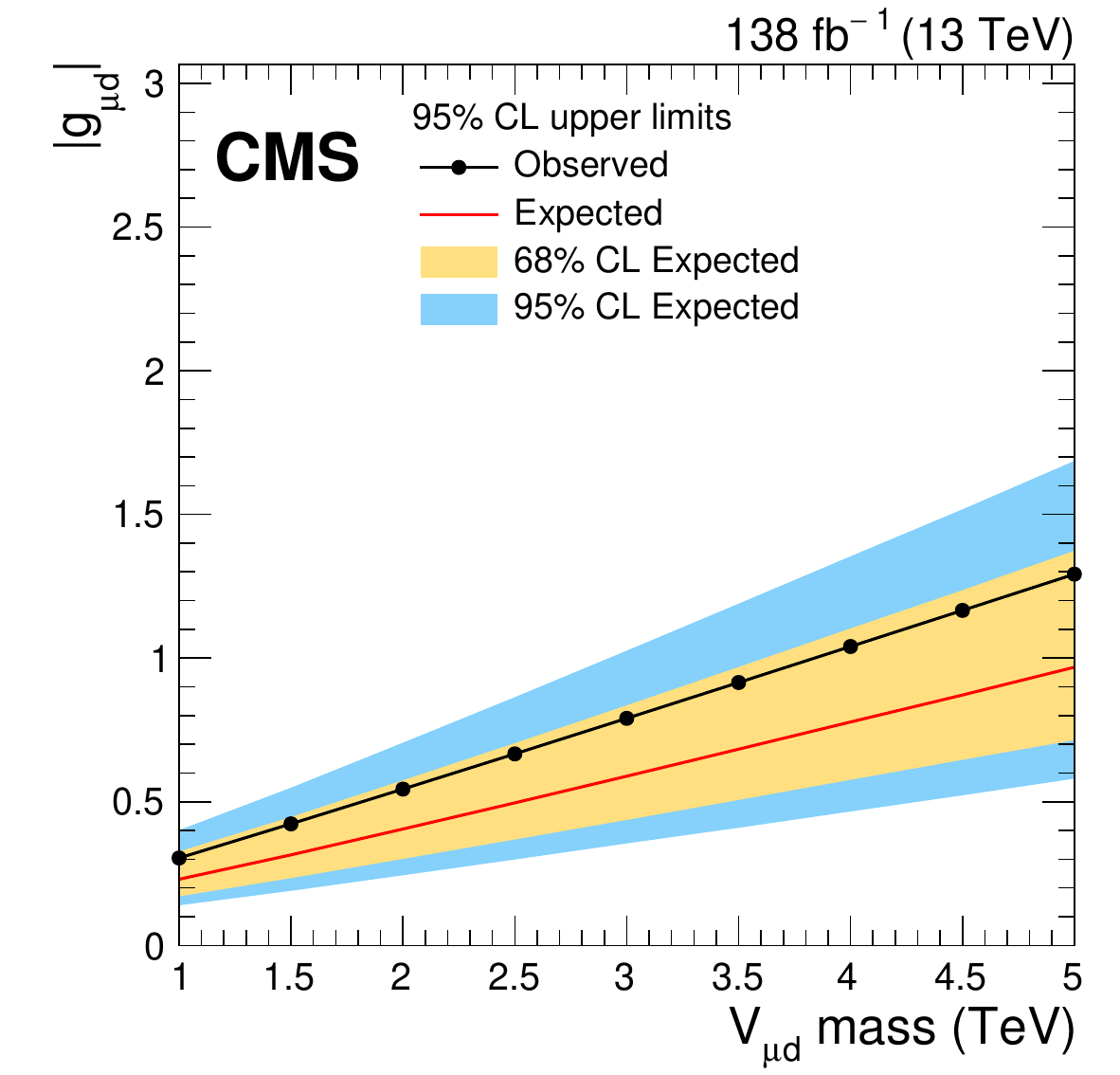}%
  }}
  \vspace{-1mm}
  \caption{
Distribution of high-\mmm dimuon data in unrolled bins of \mmm, $\abs{y}$, and \cstar (left) and exclusion limits in $\mLQ$ and the coupling strength of a vector LQ to $\text{u}\mu$ (middle) or $\text{d}\mu$ (right). Taken from Ref.~\protect\inlinecite{CMS-EXO-22-013}. %the search for nonresonant dilepton prodution in %$\LQ\to\text{d}\mu$ as a function of $\mLQ$
  }
  \label{fig:limits_nonres_dilep}
  \vspace{-2mm}
\end{figure}

% https://cms.cern.ch/iCMS/analysisadmin/cadilines?line=EXO-22-013
% https://cms-results.web.cern.ch/cms-results/public-results/publications/EXO-22-013/
% https://gitlab.cern.ch/tdr/papers/EXO-22-013
\subsection{Nonresonant dielectron \& dimuon}
As mentioned in Section~\ref{sec:overview}, nonresonant dilepton production through LQ would interfere with Drell--Yan. In particular, the angular distributions like rapidity $y$ or $\cstar=\costhetastar$ (the cosine of the scattering angle $\thetastar$ in the Collins-Soper frame) are sensitive to the nonresonant LQ signal at extremely high dilepton masses (\mll).
The search in Ref.~\inlinecite{CMS-EXO-22-013} by the CMS collaboration exploits this fact by using similar techniques as a previous precision measurement of the forward-backward asymmetry in Drell--Yan~\cite{CMS-SMP-21-002}:
% REWEIGHTING
It constructs parametrized templates by reweighting Drell--Yan simulation following analytic formulae for the differential cross section of pure Drell--Yan, plus an interference term that scales with $\lambda^2$, and a pure LQ contribution that scales with $\lambda^4$. These templates can then be used to fit data for Drell--Yan cross section parameters ($A_0$, $A_4$) and the LQ-fermion coupling strength $\lambda$.
%\begin{align} \label{eq:dilep}
%  \frac{\dd[2]\sigma}{\dd\mll\dd\cstar}
%    \propto \left[\frac{\dd[2]\sigma}{\dd\mll \dd\cstar}\right]_\mathrm{DY}
%    &+ {\color{myblue}\lambda^2}\, %\colorbox{yellow}
%       \Nint(\mll)\left(\frac{(1+\cstar)^2}{1-\cstar+\frac{2\mLQ^2}{\mll^2}}\right) \\  
%    &+ {\color{myblue}\lambda^4}\,
%        \Npure(\mll)\left(\frac{1+\cstar}{1-\cstar+\frac{2\mLQ^2}{\mll^2}}\right)^2
%\end{align}

% SIGNAL REGION
A nonresonant LQ signal is expected to contribute most at in the high-$\mll$ tail,
so the search selects dielectron or dimuon events with at least $\mll>\SI{500}{GeV}$.
The fit parameters are extracted from data binned in $\abs{y}$, $\cstar$, and $\mll$. Figure~\ref{fig:limits_nonres_dilep} shows examples for the search of a vector $\LQ\to q\mu$.
% RESULTS
No significant deviation from the SM expectation is found, and stringent exclusion limits in signal strength vs. \mLQ phase space are set for eight different LQ models and coupling scenarios: up to \SI{5}{TeV} and beyond at large couplings, as can be seen in Fig.~\ref{fig:limits_nonres_dilep}. Limits on LQ couplings to up quarks are stronger than those for down quarks due to the higher up quark content of the proton.
%Limits on muon vs. electrons very similar.

% https://twiki.cern.ch/twiki/bin/view/AtlasPublic/SearchesPublicResults
% https://atlas.web.cern.ch/Atlas/GROUPS/PHYSICS/PAPERS/EXOT-2022-42/
% https://atlas.web.cern.ch/Atlas/GROUPS/PHYSICS/PAPERS/EXOT-2022-39/
% https://cms-results.web.cern.ch/cms-results/public-results/publications/HIG-21-001/index.html
% https://indico.in2p3.fr/event/35965/timetable/?view=standard#50-search-for-lepton-flavour-v
\subsection{Nonresonant ditau}
The search for nonresonant ditau production in Ref.~\inlinecite{ATLAS-EXOT-2022-42} was presented for the first time at this conference.% 59th Rencontres de Moriond %~\cite{Pollard} (no submission)
It is based on a cross section measurement of $\text{Z}\to\tau\tau$ in association with b jets at high invariant mass. This search incorporates a detailed study of the interference effect between the nonresonant LQ process with SM Drell--Yan, which was not taken into account by previous searches~\cite{ATLAS-EXOT-2022-39,CMS-EXO-19-016}.

\begin{figure}[ht!]
  \centerline{
  \raisebox{-0.5\height}{%
    \includegraphics[width=0.33\textwidth]{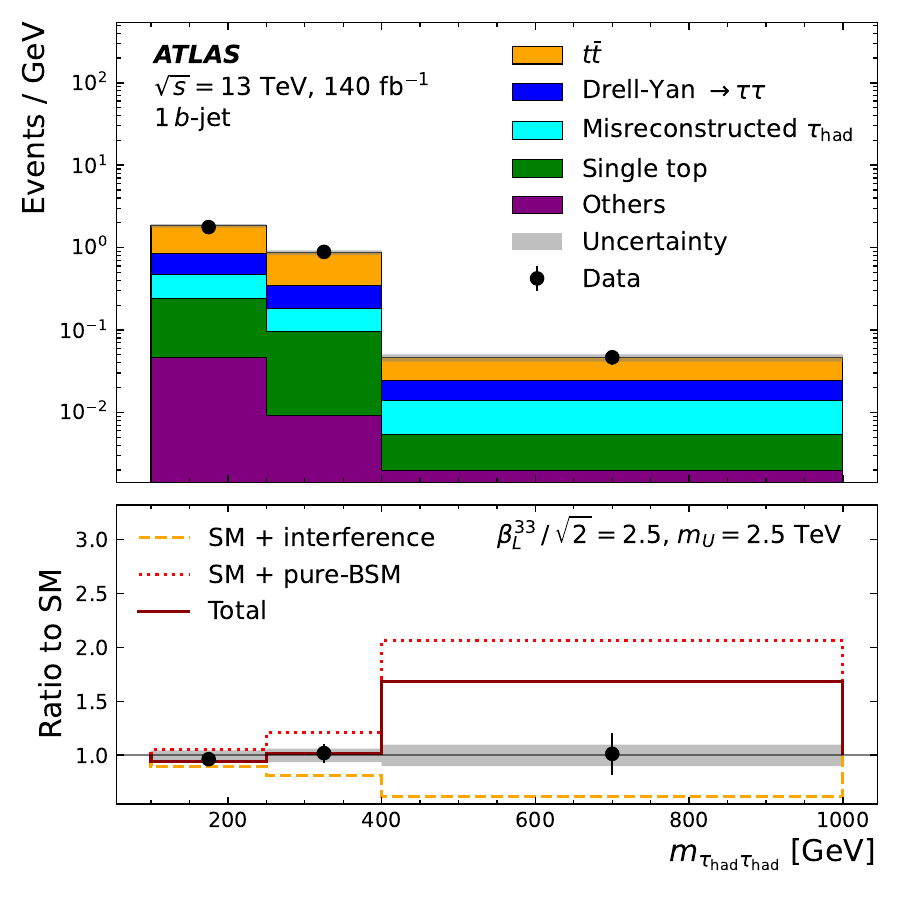}%
  }\hspace{1pt}%
  \raisebox{-0.5\height}{%
    \includegraphics[width=0.34\textwidth]{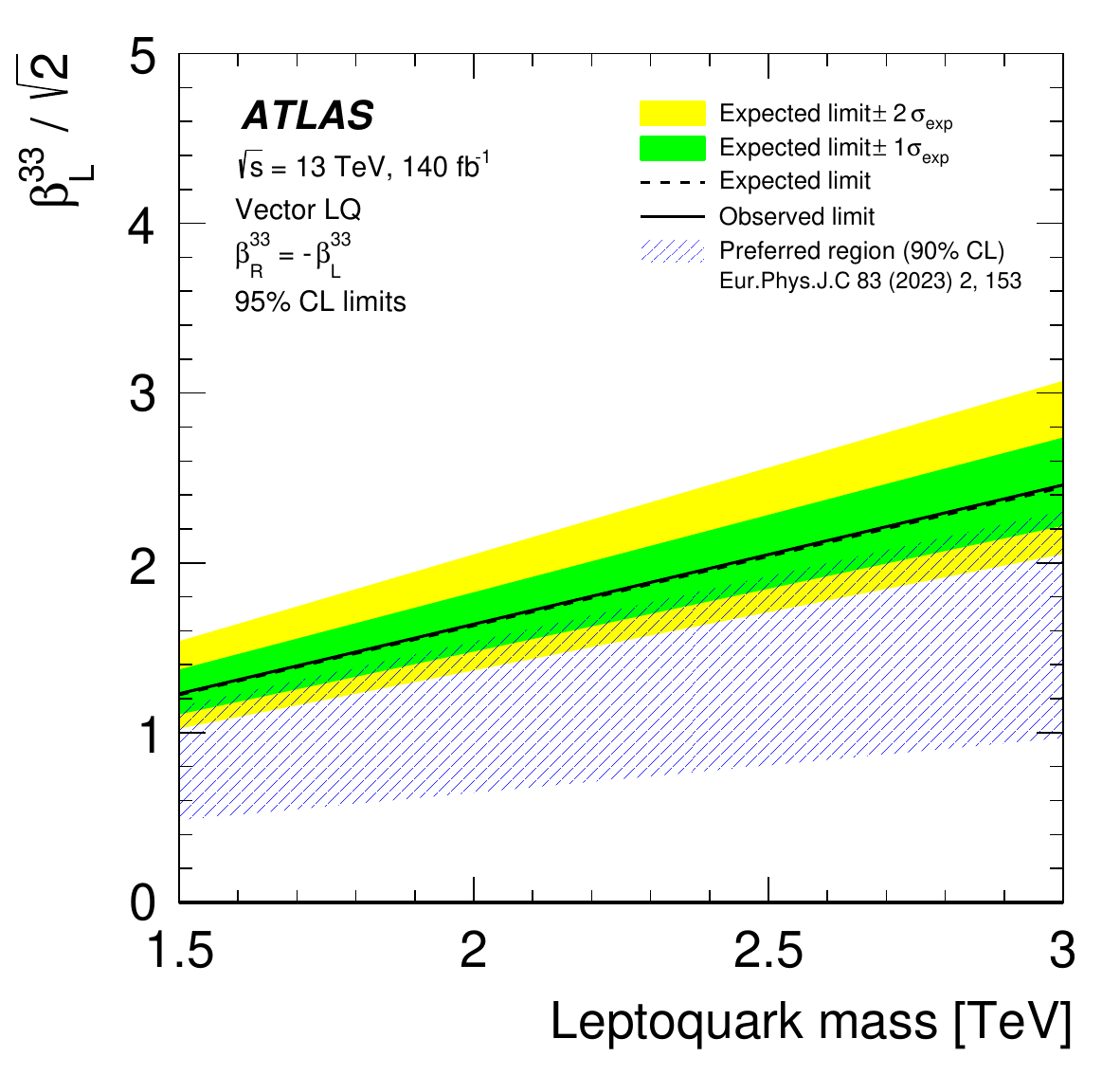}%
  }\hspace{1pt}%
  \raisebox{-0.5\height}{%
    \includegraphics[width=0.34\textwidth]{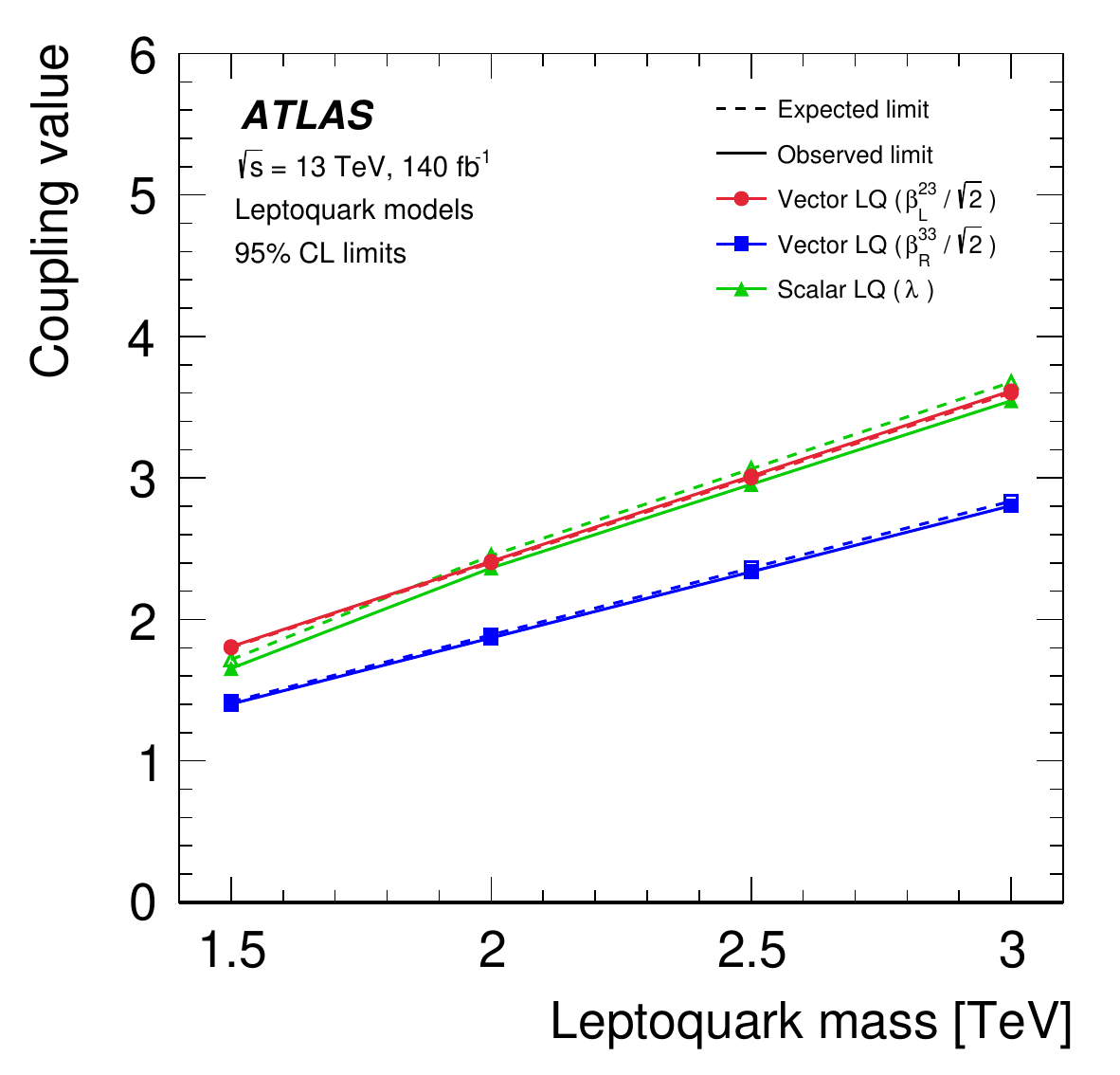}%
  }}
  \vspace{-3mm}
  \caption{
Distribution of \mthth in the 1 b-jet category (left) and exclusion limits in the $\mLQ$ and coupling strength $\lambda=\beta^{33}_L/\sqrt{2}$ phase space (middle and right).
Taken from Ref.~\protect\inlinecite{ATLAS-EXOT-2022-42}.
  }
  \label{fig:limits_nonres_ditau}
\end{figure}

% SELECTION
For $\LQ\to\PQb\tau$ couplings, nonresonant ditau production may be accompanied by b-jets, as the initial-state bottom quarks in the hard process are often produced via gluon splitting.
The search therefore selects ditau events in the fully hadronic state and categorizes the signal regions into three bins of b-jet multiplicity: zero, one, or at least two b-tagged jets. Then, a simultaneous fit is performed to the invariant mass distribution, $\mthth$, in each b-jet category. Figure~\ref{fig:limits_nonres_ditau} shows the $\mthth$ distribution in the 1 b-jet category with the expected contributions from a LQ signal.
% RESULT
No significant excess above the SM background expectation is observed, and exclusion limits in the \mLQ and signal strength phase space are set, reaching lower limits $\mLQ>\SI{3}{TeV}$ for large couplings ($\lambda > 2.4$), as shown by Fig.~\ref{fig:limits_nonres_ditau}.
With the current Run-2 data, it is possible to begin probing the phase space favored by the B anomalies (at 90\% confidence level), which is indicated by the blue hatched region in the middle plot of Fig.~\ref{fig:limits_nonres_ditau}.
This result is comparable with the search~\cite{CMS-HIG-21-001} by the CMS Collaboration, which also took into account interference. Note the latter normalizes the coupling constant as $\lambda=g_U/\sqrt{2}$.

% FIGURE: Nonres. dilep. limits (CMS-EXO-22-013)
% https://cms-results.web.cern.ch/cms-results/public-results/publications/EXO-19-016/

%%%%%%%%%%%%%%%%%%%
%   CONCLUSIONS   %
%%%%%%%%%%%%%%%%%%%

\section{Conclusions}

% MOTIVATION
The leptoquark (LQ) is a new type of hypothetical boson that is well motivated by many theories beyond the Standard Model, and recent experimental anomalies.
% EXPERIMENT
Searches for LQs in proton-proton collisions at the ATLAS and CMS experiments have explored many final states in the Run-2 dataset collected between 2016--2018, including signatures for the pair or single production of a LQ, or nonresonant dilepton production through the $t$-channel exchange of a LQ.
Although many well-motivated final states have been covered, many opportunities remain, such as signatures containing an electron or muon accompanied by a light- or charm-flavored jets.
Unfortunately, no direct evidence for LQ has been observed, and stringent exclusion limits in terms of LQ mass, coupling strength to fermions, and branching fractions have been set.
% REACH
Searches for pair production can now reach up to about \SI{2}{TeV} or beyond,
while recent searches for a nonresonant LQ signal open a large phase space up beyond an impressive \SI{5}{TeV} at high coupling strengths.

%%%%%%%%%%%%%%%%%%
%   REFERENCES   %
%%%%%%%%%%%%%%%%%%

\section*{References}
\bibliography{INeutelings}

\end{document}